\documentclass[12pt,a4paper,final]{iopart}

\usepackage{iopams}  
\usepackage{graphicx}
\usepackage[breaklinks=true,colorlinks=true,linkcolor=blue,urlcolor=black,citecolor=blue]{hyperref}

\usepackage{tabularx}
\usepackage{color}
\usepackage{subcaption}
\usepackage{float}
\usepackage{cite}

\usepackage{eqnarray}

\begin{document}
\title[]{Characterisation of a charged particle levitated nano-oscillator}

\author{N. P. Bullier}%
\ead{nathanael.bullier.15@ucl.ac.uk}
\address{Department of Physics and Astronomy, University College London, Gower Street, London WC1E 6BT, United Kingdom}%

\author{A. Pontin}
\ead{a.pontin@ucl.ac.uk}
\address{Department of Physics and Astronomy, University College London, Gower Street, London WC1E 6BT, United Kingdom}

\author{P. F. Barker}
\ead{p.barker@ucl.ac.uk}
\address{Department of Physics and Astronomy, University College London, Gower Street, London WC1E 6BT, United Kingdom}%
\begin{abstract}
We describe the construction and characterisation of a nano-oscillator formed by a Paul trap. The frequency and temperature stability of the nano-oscillator was measured over several days allowing us to identify the major sources of trap and environmental fluctuations. We measure an overall frequency stability of 2\,ppm/hr and a temperature stability of more than 5\,hours via the Allan deviation. Importantly, we find that the charge on the nanoscillator is stable over a timescale of at least two weeks and that the mass of the oscillator, can be measured with a 3\,\% uncertainty. This allows us to distinguish between the trapping of a single nanosphere and a nano-dumbbell formed by a cluster of two nanospheres.

\end{abstract}


\maketitle
\section{Introduction}

Nanoscale levitated optomechanical systems are currently of considerable interest for exploring the macroscopic limits of quantum mechanics \cite{Oriol,Chang,Peter}, for sensing of short range forces \cite{Geraci2}, and for exploring non-equilibrium thermodynamics \cite{Rondin2017}. Unlike typical optomechanical systems, levitation offers good isolation from the environment while only a few mechanical degrees of freedom need be considered. The ability to turn-off the trapping potential allows field free experiments such as matter-wave interferometry \cite{Kaltenbaek2016,Bateman2014} and precision measurements of forces \cite{Ranjit}. Optical and magnetic levitation of nanoparticles have been demonstrated with cooling to microKelvin temperatures \cite{Slezak2018,Felix2019}. While optical levitation is well developed and allows relatively high mechanical frequencies, it is fundamentally limited for many applications by the unavoidable photon recoil heating introduced by the laser used for trapping \cite{Jain2016}. Levitation of charged nanoparticles via a Paul trap, is an attractive alternative that minimises or even avoids such heating. A hybrid electro-optical cavity system has been used to demonstrate cavity cooling of nanoparticles \cite{Millen2015,Fonseca2016}. Similar schemes are being considered for testing mechanisms of wavefunction collapse at macroscopic scales by measuring the reheating rate of a cooled oscillator \cite{Goldwater,TEQ}. In order to perform both of those experiments, the stability and the noise needs to be well characterised. Moreover, a precise knowledge of the mass, the number of charges and their stability over time is required. For instance in the hybrid trap, the cooling process depends on the number of charges. The Paul trap is a promising platform for these studies as it has been a key tool for quantum science and technology \cite{electron,Paul,Monroe}. It has been utilised for the creation of stable atomic clocks and for the demonstration of important protocols in quantum computation and information \cite{clock,Monroe}. In these traps, atomic and molecular ions can be laser cooled to their ground state and trapped in isolation for days. Key to their utilisation has been that a deep and stable low noise electrical potential can be readily created. Many charged nanoparticle traps \cite{Schlemmer,Cai,Monroe,Millen2015,Alda,Kane,Bullier,Heather,Tracy} have been demonstrated but there are few reports characterising their long term stability and noise, which is crucial for applications in quantum optomechanics and for testing fundamental physics.

In this paper we describe the detailed characterisation of a nano-oscillator formed by a Paul trap and assess this platform for its use in quantum optomechanics experiments including a hybrid electro-optical trap \cite{Millen2015,Fonseca2016} and for future non-interferometric tests of the macroscopic limits of quantum mechanics \cite{Goldwater,TEQ}. We discuss different ways of measuring the size and mass of the particle and estimate its centre-of-mass temperature based on those results. We describe measurements that determine the stability of the Paul trap over time, and the effects of the loading mechanism and temperature fluctuations.

\bigskip
\captionsetup[subfigure]{labelformat=empty}

\section{Paul trap description}

\begin{figure}[!ht]
\centering
\includegraphics[width=1\linewidth]{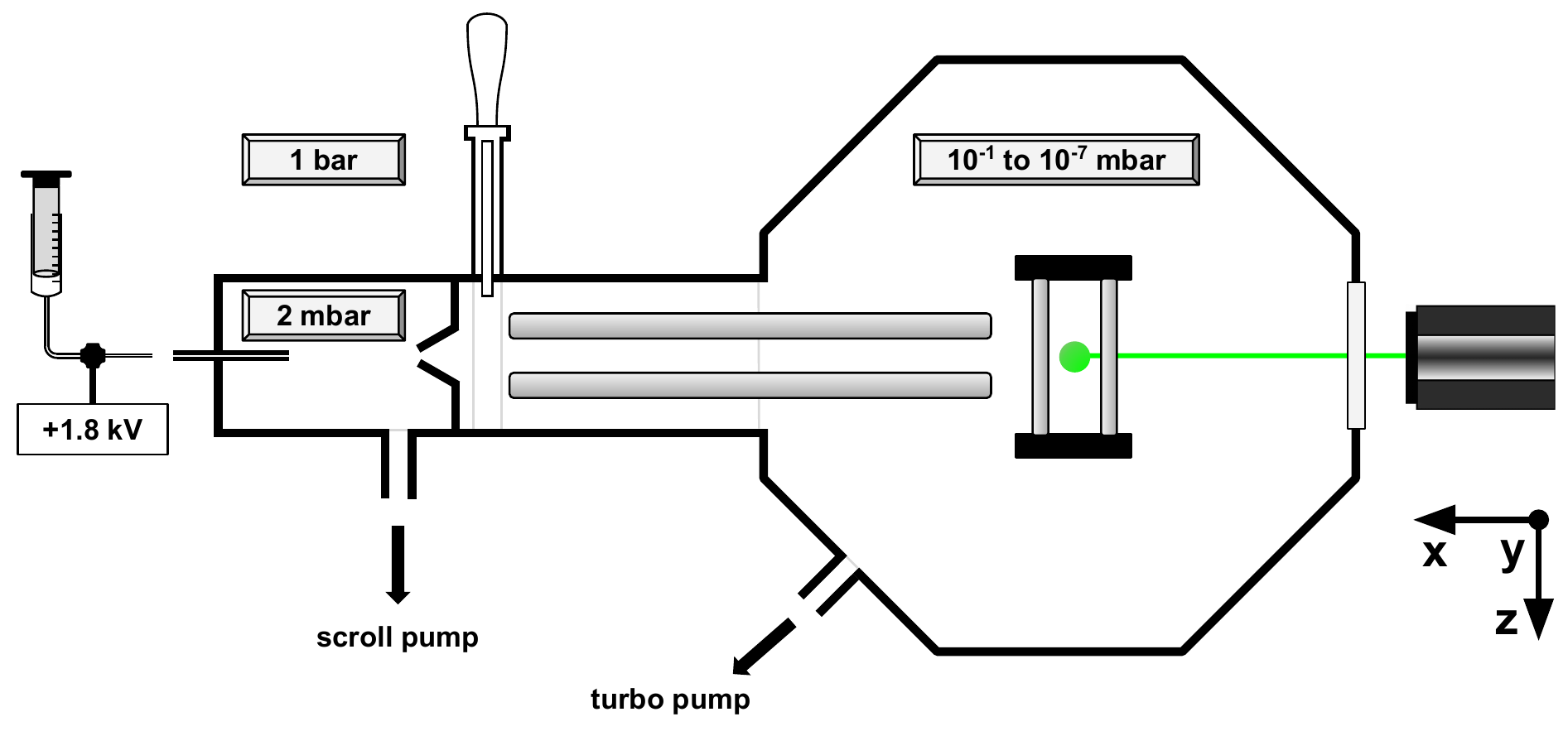}
\caption{A schematic of the experimental setup. Nanoparticles suspended in ethanol are electrosprayed at atmospheric pressure. The aerosol is entrained into the first pumping stage which is kept at a pressure of $2$\,mbar. The charged nanoparticles are directed through a beam skimmer and enter the main vacuum chamber, which is kept at a pressure of $\sim10^{-1}$\,mbar during the loading phase. It is then guided inside an electric quadrupole guide towards the trap. The nanoparticle is illuminated with a green diode laser along the \textit{x}-axis. The motion of the particle is imaged in the \textit{x-z} plane with a low-cost CMOS camera placed along the \textit{y}-axis.}
\label{fig0}
\end{figure}

\begin{figure}[!ht]
\centering
\includegraphics[width=.97\linewidth]{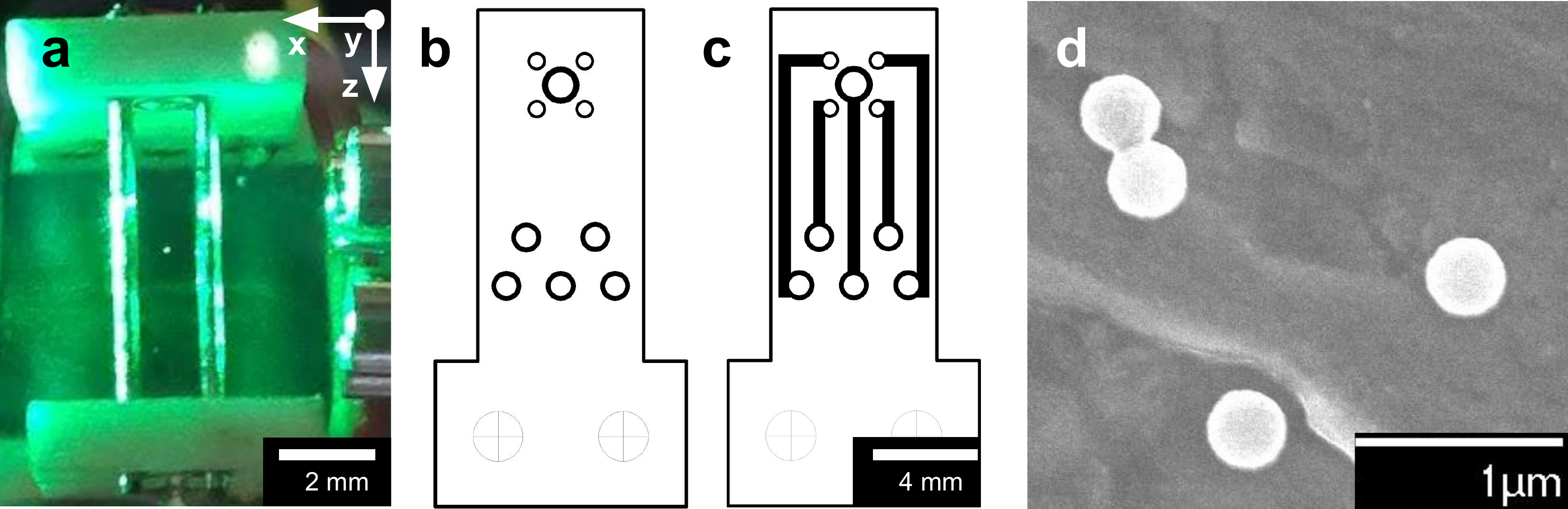}
\caption{(a) An image of the linear Paul trap loaded with a nanoparticle in its centre. The end of the quadrupole guide can be seen on the right-hand side of the picture. (b) The PCB layout of the internal side of the trap. The five holes at the top of the layout are used for the trap connections. The pad in the centre of the five holes is used as an endcap. The other four holes are used as both holders and electrical connections to the four rods constituting the trap. Tracks made out of gold coated copper are shown in black. (c) The PCB layout on the external side of the trap. The connections to the PCB board are done on the lower part of the PCB on the five lower holes. (d) A scanning electron microscope (SEM) image of the nanoparticles used. The trap parameters discussed in the main text are the following: $r_o=1.1$\,mm, $z_o=3.5$\,mm, $\eta=0.82$, $\kappa=0.086$, and $\epsilon=0.5$.}
\label{fig1}
\end{figure}

A levitated nano-oscillator is produced by trapping a single charged silica nanoparticle in a Paul trap. Silica nanopsheres are suspended in ethanol at a concentration of 10\,$\mu$g/mL. Prior to loading, the solution is left in an ultrasonic bath for 20\,minutes to restore monodispersion. As shown in Fig.\,\ref{fig0}, an electrosol of this solution is created by means of electrospray ionisation \cite{electrospray,Kane} by applying 1.8\,kV (of either polarity) on a needle with a 100\,$\mu$m internal diameter. This aerosol of particles are entrained into the gas flow which enters a capillary tube that leads directly into the first pumping stage. The capillary is 5\,cm long with a 250\,$\mu$m internal diameter. This first vacuum stage is kept at a pressure of approximately 2\,mbar  by directly connecting it to a scroll pump. The flux of particles travels thorugh a skimmer with a 0.4\,mm aperture. The skimmer isolates the first stage from the main chamber which is initially  kept at a pressure of $\sim 10^{-1}$\,mbar during the loading phase. Once in the main chamber, some of particles are guided through a 20\,cm long quadrupole guide and their motion is damped by collisions with the surrounding gas so that the particles have low enough energy to be captured by the Paul trap. The guide increases the flux of particles reaching the trap and is also shown in Fig.\,\ref{fig1}\,(a). Its end is placed $\sim 3\,$mm away from the Paul trap. By operating the guide in a mass filter configuration \cite{march}, where a high DC voltage is applied onto two electrodes, one can select a given charge-to-mass ratio. Note that since the trap and the guide have different geometries, the effective stability region is further reduced. We typically trap particles with charge-to-mass ratios in the range $0.05<q/m<2$\,C/kg. Once the trap is loaded, the guide is grounded to avoid any excess micromotion caused by the AC field of the guide. While loading the trap, droplets of solvent can be trapped as well as nanoparticles and can remain for hours in low vacuum. In order to ensure that we only observe bare silica nanoparticles, the pressure is reduced to $\sim 10^{-4}$\,mbar right after trapping to increase the evaporation speed. Currently, the pressure can then be reduced in the main chamber down to $\sim10^{-7}$\,mbar and the particles can remain trapped at those pressures for weeks without any cooling. 

An image of the linear Paul trap used for these studies is shown in Fig.\,\ref{fig1}\,(a). An AC potential is applied onto two rods diagonally opposed, while the other two are grounded. The effective potential confines a charged particle in the \textit{x-y} plane. An additional DC voltage is applied onto two endcap electrodes to trap the particle along the \textit{z-}axis. With this geometry, the electric potential $\Phi(x,y,z,t)$ close to the centre of the trap is given by \cite{wineland}

\begin{eqnarray}
\label{eq1}
\hspace{-15mm}
\Phi(x,y,z,t)=\frac{V_o}{2} \left(1+\eta\frac{x^{2}-y^{2}}{r_o^{2}} \right) \cos(\omega_{d}t)+\frac{\kappa\, U_o}{z_o^2}\left(z^{2}-\epsilon x^{2}-(1-\epsilon) y^{2} \right),
\end{eqnarray}

\noindent where $r_o$ is the distance between the centre of the trap and the rod electrodes providing the AC field. The distance between the centre of the trap and the endcap electrodes which provide the DC field is defined to be $z_o$. $V_o\,\cos(\omega_{d}t)$ and $U_o$ are the applied potentials, $\omega_{d}$ is the drive frequency, and $\eta$ and $\kappa$ are geometrical efficiency coefficients \cite{efficiency1,efficiency2} quantifying non-perfect quadratic potentials. The trap ellipticity is given by $\epsilon$, which is here equal to 0.5 because of the symmetry of the trap in the \textit{x-y} plane. Those parameters are obtained from numerical finite element method (FEM) simulations of the electric field with values given in the caption of Fig.\,\ref{fig1}. 

The trapped particle feels an effective pseudo-potential $\phi(x,y)=\frac{q}{4m\omega_{d}^{2}}|\nabla V(x,y)|^2$ with $V(x,y)=\frac{V_{0}}{2}\left(1+\eta\frac{x^{2}-y^{2}}{r_o^{2}} \right)$, where $q/m$ is the charge-to-mass ratio of the nanoparticle. The dynamics of the nanosphere in the potential shown in Eq.\,\ref{eq1} are governed by three Mathieu equations with stability parameters $a_{i}$ and $q_{i}$ \cite{wineland}. They quantify the stability of the motion along the \textit{i-}axis given by the AC and DC potential, respectively and are given by

\begin{equation}
\hspace{20mm}
\eqalign{
\label{eq2}
a_{x}=a_{y}=-\frac{1}{2}a_{z}=-\frac{q}{m}\frac{4\kappa\,U_{o}}{z_{o}^{2}\,\omega_{d}^{2}}\,, \\\cr
q_{x}=-q_{y}=\frac{q}{m}\frac{2\eta V_{o}}{r_{o}^{2}\,\omega_{d}^{2}},\,q_{z}=0\,.\cr}
\end{equation}

\vspace{20mm}
\noindent If $a_{i}\ll1$ and $q_{i}\ll1$, the secular frequencies can be well approximated by

\begin{eqnarray}
\label{eq3}
\hspace{30mm}
\omega_{i}\approx\frac{\omega_{d}}{2}\sqrt{a_{i}+\frac{1}{2}q_{i}^{2}}\,.
\end{eqnarray}

The electrode supports are realised on a printed circuit board (PCB) \cite{PCB} whose layout is shown in Fig.\,\ref{fig1} (b) and (c). The stainless steel electrodes of the trap are rods of 0.5\,mm diameter which seat into the holes of the PCB. One plated through hole on each PCB in the middle of the rods is used as an endcap. The two PCB boards (and therefore the endcaps) are kept at a distance of 7.0\,mm. Both the traces and the holes are gold coated to minimise patch potentials \cite{patch}. Using a PCB to support the electrodes allows flexibility in making the electrical connections to the rods and keep to a minimum their effect on the potential of the trap. The substrate used here (RO4003C \cite{PCB}) is a ceramic-filled woven glass that is compatible with ultra-high vacuum \cite{Rouki_2003,Westerberg2003}. The PCB tracks are separated at least by 225\,$\mu$m and are chosen to avoid a voltage breakdown at any pressure with the typical AC voltages applied of $\sim$\,500\,V peak-to-peak.

\hspace{10mm}
\section{Particle detection and characterisation}
\label{sec:particle}

The particle is detected by illumination with a green laser diode directed along the \textit{x-}axis of the trap. A typical power of 40\,mW with a beam waist of 250\,$\mu$m was used. The motion of the particle in the \textit{x-z} plane is monitored by using a low-cost CMOS camera which detects the scattered light in the \textit{y-}direction (see Fig.\,\ref{fig0}) \cite{camera_paper}. This enables us to detect the stochastic motion of the nanoparticle driven by thermal noise and other unknown noises from which we can extract the particle damping $\gamma$, frequency of oscillation $\omega_{i}$ (along the \textit{i$^{th}$}-axis) and amplitude. The one-sided power spectral density (PSD) of the motion along the \textit{i$^{th}$}-axis can be written as $S_{i}(\omega)=|\chi(\omega)|^{2}(S_{th}+S_{F}(\omega))$ where $\chi(\omega)= (m(\omega_{i}^{2}-\omega^{2}-i\gamma\omega))^{-1}$ is the mechanical susceptibility. The thermal force noise is given by $S_{th}=4k_{B}Tm\gamma$ where $m$ is the particle mass, $T$ the bath temperature, $k_{B}$ the Boltzmann constant, and $S_{F}(\omega)$ the force noise of unknown sources. We use commercial silica nanospheres by \textit{microParticles GmbH} with nominal density of 1850\,kg/m$^{3}$ and nominal radius of (194\,$\pm$\,5)\,nm. An SEM picture of the particles used is shown in Fig.\,\ref{fig1} (d). We find a radius of (186\,$\pm$\,4)\,nm from a sample of 223 imaged nanoparticles.

 \begin{figure}[!ht]
\centering
\begin{subfigure}{.5\textwidth}
  \centering
  \includegraphics[width=.97\linewidth]{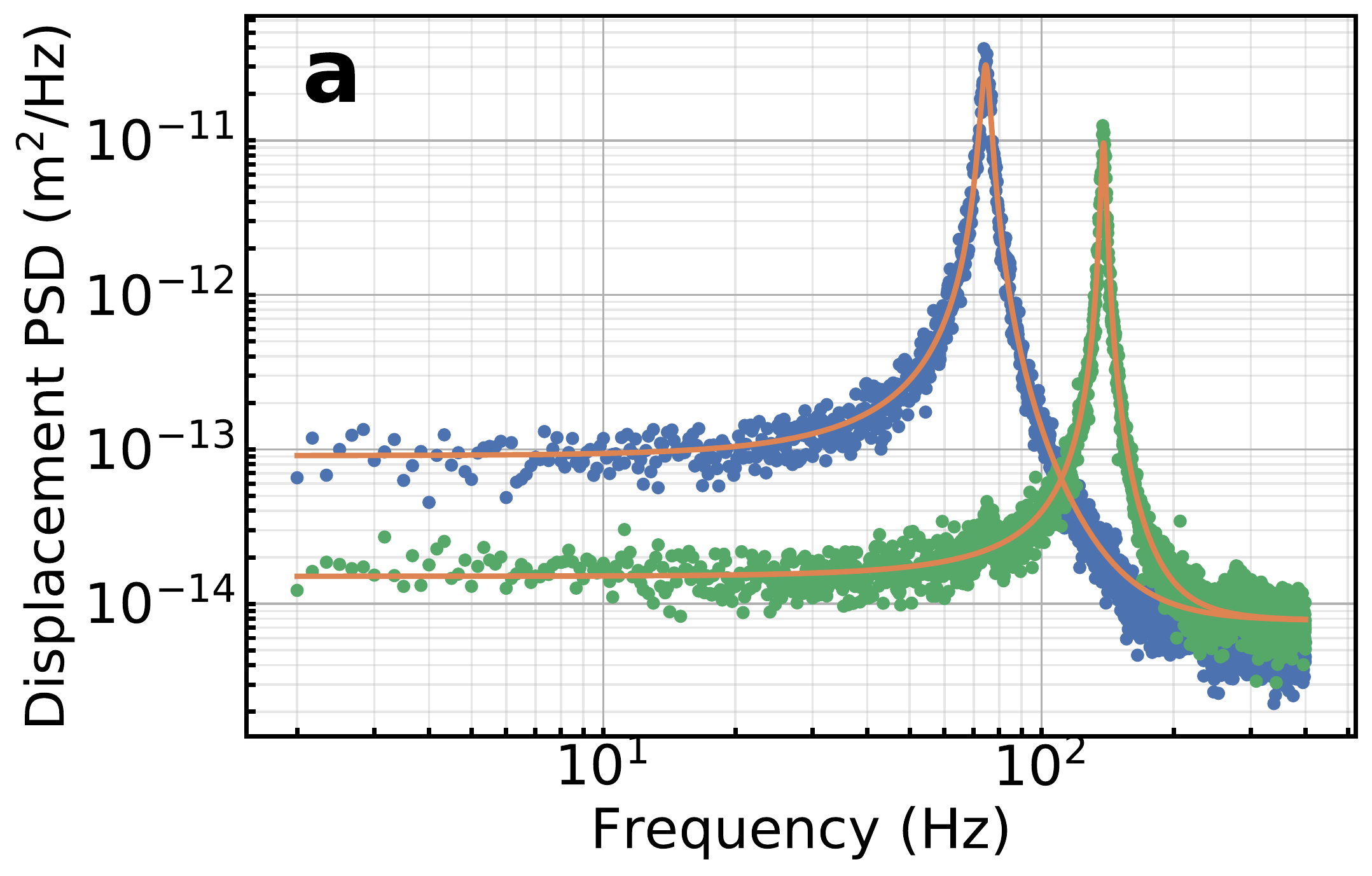}
  \caption{}
  \label{fig:cal1}
\end{subfigure}%
\begin{subfigure}{.5\textwidth}
  \centering
  \includegraphics[width=1\linewidth]{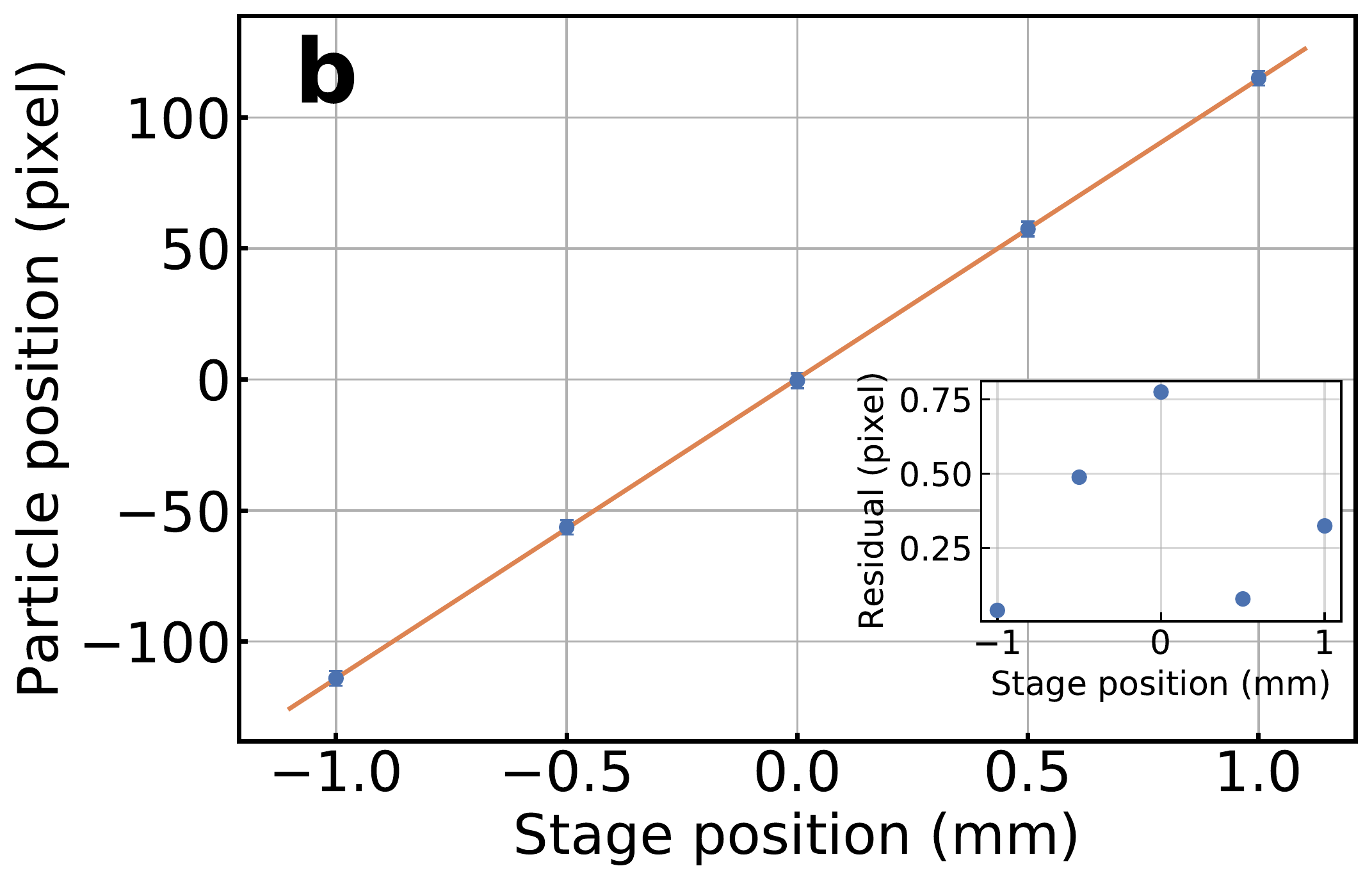}
  \caption{}
  \label{fig:cal2}
\end{subfigure}
\caption{(a) Calibrated displacement spectra of a levitated silica nanopshere at 1.2$\times$10$^{-2}$\,mbar. The figure shows the displacement PSD in blue (green) along the \textit{z-}axis (\textit{x}-axis). A fit to the PSD is shown in orange from which are extracted the secular frequency and the mechanical damping. (b) The camera is calibrated by moving it by a known amount. The figure shows the sum of the relative distances of the translation stage positions. The fit is shown in orange with the residuals in the inset. The calibration factor is found to be (8.75$\,\pm\,0.09$)\,$\mu$m/pixel.}
\label{fig:cal}
\end{figure}

\subsection{Calibration of the motion}
We calibrate the motion of the particle by placing the camera on a translation stage and displacing it by a known amount. Five time traces are taken at different stage positions. By calculating the mean position of the particle for each time trace, we directly map the stage position onto the camera pixel matrix since the particle behaves like a point source. After a simple linear regression, the example shown in Fig.\,\ref{fig:cal} (b) gives a displacement per pixel of (8.75$\,\pm\,0.09$)\,$\mu$m/pixel. Two independent uncertainties are taken into account. The uncertainty of the fit itself, which only accounts for 0.3\%, while the uncertainty in the camera focus is 1\%. When out of focus, the mean position of the particle remains the same but the standard deviation increases. This error is estimated by changing the focus on the trapped particle in a controlled manner. The focus is checked over days to correct for drifts in the camera mount or focus. Lastly, by ensuring that the residuals of the fit do not increase when the camera is displaced away from the nanoparticle, we ensure that distortions due to the camera objective do not have to be taken into account. This method is competitive for measurements of low frequency oscillators in comparison to methods measuring the particle response to a known force, which can depend on more parameters with larger associated uncertainties \cite{Hebestreit2018}. We show an example of a calibrated displacement PSD of the nanoparticle in Fig.\,\ref{fig:cal} (a).

\subsection{Size estimation}
\begin{figure}[H]
\centering
\begin{subfigure}{.5\textwidth}
  \centering
  \includegraphics[width=0.97\linewidth]{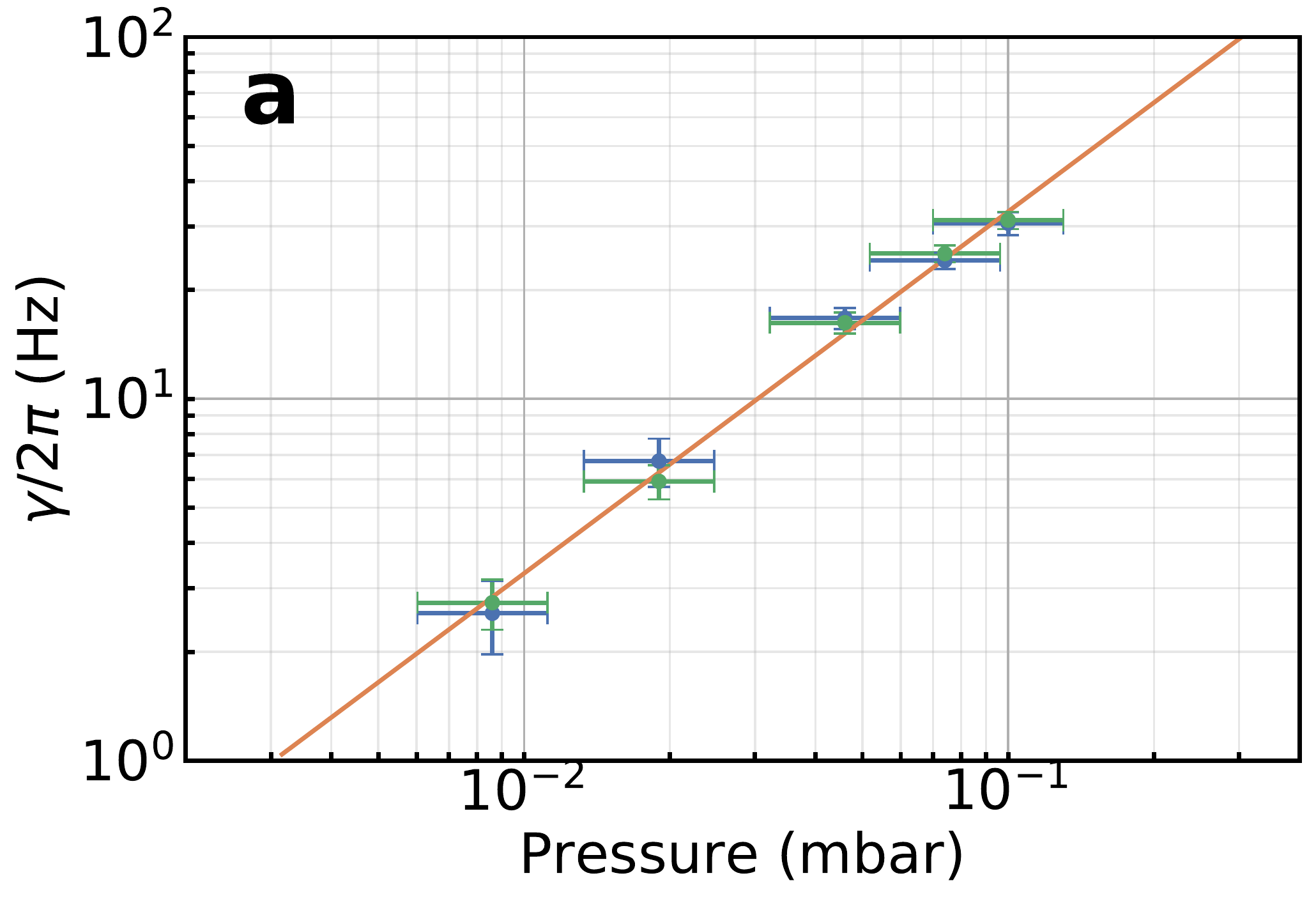}
  \caption{}
  \label{fig:sub1}
\end{subfigure}%
\begin{subfigure}{.5\textwidth}
  \centering
  \includegraphics[width=1\linewidth]{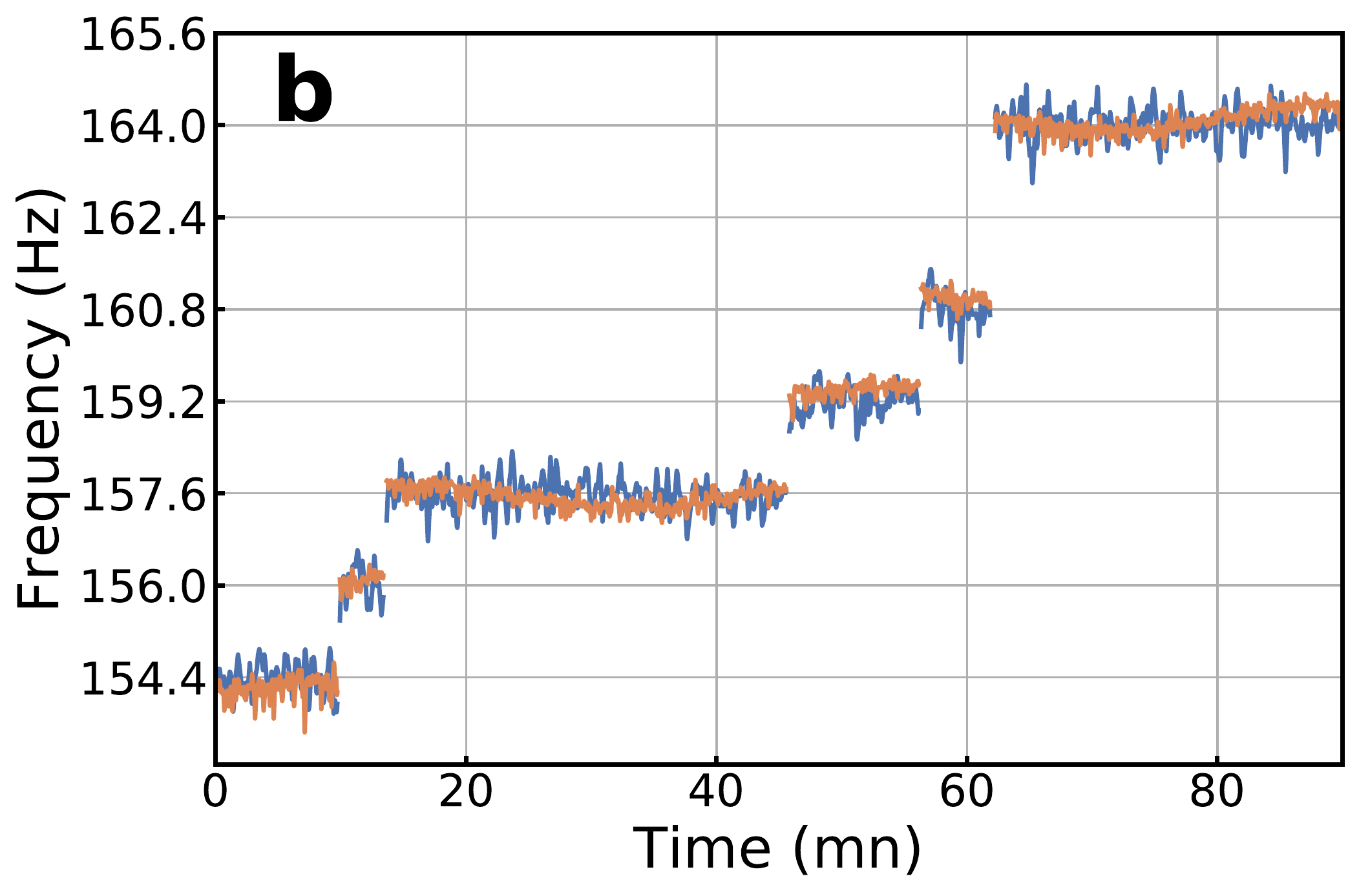}
  \caption{}
  \label{fig:sub2}
\end{subfigure}
\caption{Estimation of the particle size. (a) Gas damping as a function of pressure. Linewidth measurements from spectra are shown in blue and green corresponding respectively to the damping along the \textit{x} and \textit{z-}axis. The fit is shown in orange. The radius found is (199\,$\pm$\,57)\,nm assuming a density of 1850\,kg/m$^{3}$ and a spherical shape. (b) Monitoring of the secular frequencies over time along the \textit{x} and \textit{z}-axis (\textit{z}-axis rescaled) shown respectively in orange and blue. A grid with constant separation of 1.59\,$\pm$\,0.07\,Hz is added to the plots. It corresponds to the frequency shift given by one elementary charge. From the frequency shifts, we estimate the mass to be (9.6$\,\pm\,0.9$)$\times$10$^{-17}$\,kg.}
\label{fig:size}
\end{figure}

Different methods can be used to estimate the size (or mass) of the nanoparticles loaded into the trap with different accuracies \cite{Li2011,Slezak2018,Schlemmer,Sung,Cai,Ricci2019}. The particle size can be roughly determined by measuring the gas damping as a function of pressure. This is done by measuring the linewidth of the displacement PSD at different pressures. It is then fitted to the expected gas damping law in the free molecular flow regime. Here \cite{damping,damping2}:$\gamma_{\mathrm{gas}}=\frac{(1+\frac{\pi}{8})\bar{c}P_{g}\,m_{g}}{k_{B}\,T_{b}\,a\rho}$, where the particle mean speed is $\bar{c}=\sqrt{8k_{B}T_{b}/m_{g}/\pi}$, $a$ and $\rho$ the particle radius and density, respectively, $k_{B}$ the Boltzmann constant, $P_{g}$, $m_{g}$ and $T_{b}$ correspond to gas pressure, the mass of the gas molecules and the bath temperature, respectively. The bath temperature $T_{b}$ can be assumed to remain at 293\,K independently from both the centre-of-mass motion and the internal temperature of the nanosphere \cite{Millen2014}. This gives a radius of (199$\,\pm\,57$)\,nm by assuming a nominal density of 1850\,kg/m$^{3}$ and a spherical shape (see Fig.\,\ref{fig:size} (a)). The main uncertainty contribution is due to the pressure measurement (30\%, specified by the manufacturer). On top of its large uncertainty, this method has a few drawbacks. It requires the knowledge of the density, which can significantly vary for silica nanospheres \cite{Parnell2016}. Furthermore, it only works for spheres and it can therefore be challenging to differentiate a single sphere from a cluster. Indeed, in the case of two particles joined together such as a nanodumbbell, the expected linewidth should be smaller by 8\,\% compared to a sphere (independently from the size of the spheres and assuming a stochastic rotational motion of the dumbbell) \cite{damping}, which is comparable to the statistical uncertainty of the measurement shown here. In comparison, a change in mass by a factor 2 for a spherical object corresponds to a reduction in linewidth by 21\,\%, easier to measure. It is therefore challenging to differentiate with this method a nanodumbbell from a single nanosphere in a Paul trap. It can be more easily differentiated, when the alignment with respect to the trap axis is well defined, such as in an optical tweezer \cite{nanodumbbell}.

In a Paul trap, a more attractive method to determine the mass consists of directly measuring it from the secular frequency shifts due to charge jumps \cite{Schlemmer,Sung}. For a number of charges larger than $\sim$\,50, the secular frequency along the \textit{x}-axis can be approximated as $\omega_{x}\approx\frac{\omega_{d}}{2\sqrt{2}}q_{x}$ (see Eq.\,\ref{eq2} and \ref{eq3}). The frequency shift given by one electron becomes $\delta\omega_x=\frac{|e|\,V_{0}\,\eta}{\sqrt{2}m\,\omega_{d}\,r_{o}^{2}}$  where $e$ is the elementary charge. While we have monitored a constant number of charges for days, it can be changed (mostly increased in our case) by leaving the pressure gauge on \cite{frimmer}, which gives a typical charging rate of $\sim$\,7\,charges/hour. Other ways of charging nanoparticles include using an electron gun \cite{Sung} or a UV light source \cite{Slezak2018}. The jumps in the secular frequency can be seen in Fig.\,\ref{fig:size} (b). Assuming that the smallest frequency jump corresponds to a change in a single elementary charge, we measure a mass of (9.6$\,\pm\,0.9)\times10^{-17}$\,kg. Two strong motivations to use this method are that thermal equilibrium of the centre-of-mass motion with the bath is not required and no knowledge over the density is required. The systematic error in the mass uncertainty comes from the trap fabrication tolerances of $\pm\,50\,\mu$m in both the relative position of the holes and their diameter. This gives an uncertainty of 4\,\% on $r_o$. The statistical error accounts for 4\,\% and therefore the overall error on the mass is 9\,\%. Using this method, statistical uncertainties as small as 100\,ppm with systematic uncertainty of 1\,\% have been demonstrated in mass-spectrometers \cite{Schlemmer,Sung}. 

The mass of the particle can also be estimated by assuming thermal equilibrium at 293\,K with the bath. We verify in the next section the validity of this assumption. However, it is reasonable because the very low intensity (40\,W/cm$^{2}$ against 10\,MW/cm$^{2}$ for a typical optical tweezer) used to illuminate the particle is not sufficient to increase its internal temperature \cite{Millen2014}. Moreover, these measurements are taken at a high pressure ($10^{-2}$\,mbar) where the heat transfer to the surrounding gas is very efficient. At this pressure, thermal noise dominates other sources of noise e.g. electrical noise. Following from the equipartition theorem, at thermal equilibrium, $\frac{1}{2}k_{B}T_{b}=\frac{1}{2}m\omega_{i}^{2}\langle i \rangle ^{2}$, where $\langle i\rangle$ is the standard deviation of the motion along the \textit{i$^{\textit{th}}$-}axis. This measurement offers the smallest uncertainty over the different mass measurements used here (3\,\%) with $m=(9.5\,\pm\,0.3)\times10^{-17}$\,kg. The mass was obtained after averaging 16 measurements. The total error is composed of a $1\,\%$ statistical error and a $3\,\%$ systematic error that takes into account the temperature uncertainty of $2\,\%$, an error of $1\,\%$ on the frequency and of $2\,\%$ on the variance of the displacement given by the calibration method. The uncertainty on this measurement could easily be reduced down to 2\,\% by using a more precise temperature sensor. Another method, combining the thermal equilibrium assumption with a known force excitation has been demonstrated in Ref.\,\cite{Ricci2019}. It would however require a precise knowledge of the number of charges to be applied here.

\begin{figure}[!ht]
\centering
\includegraphics[width=.7\linewidth]{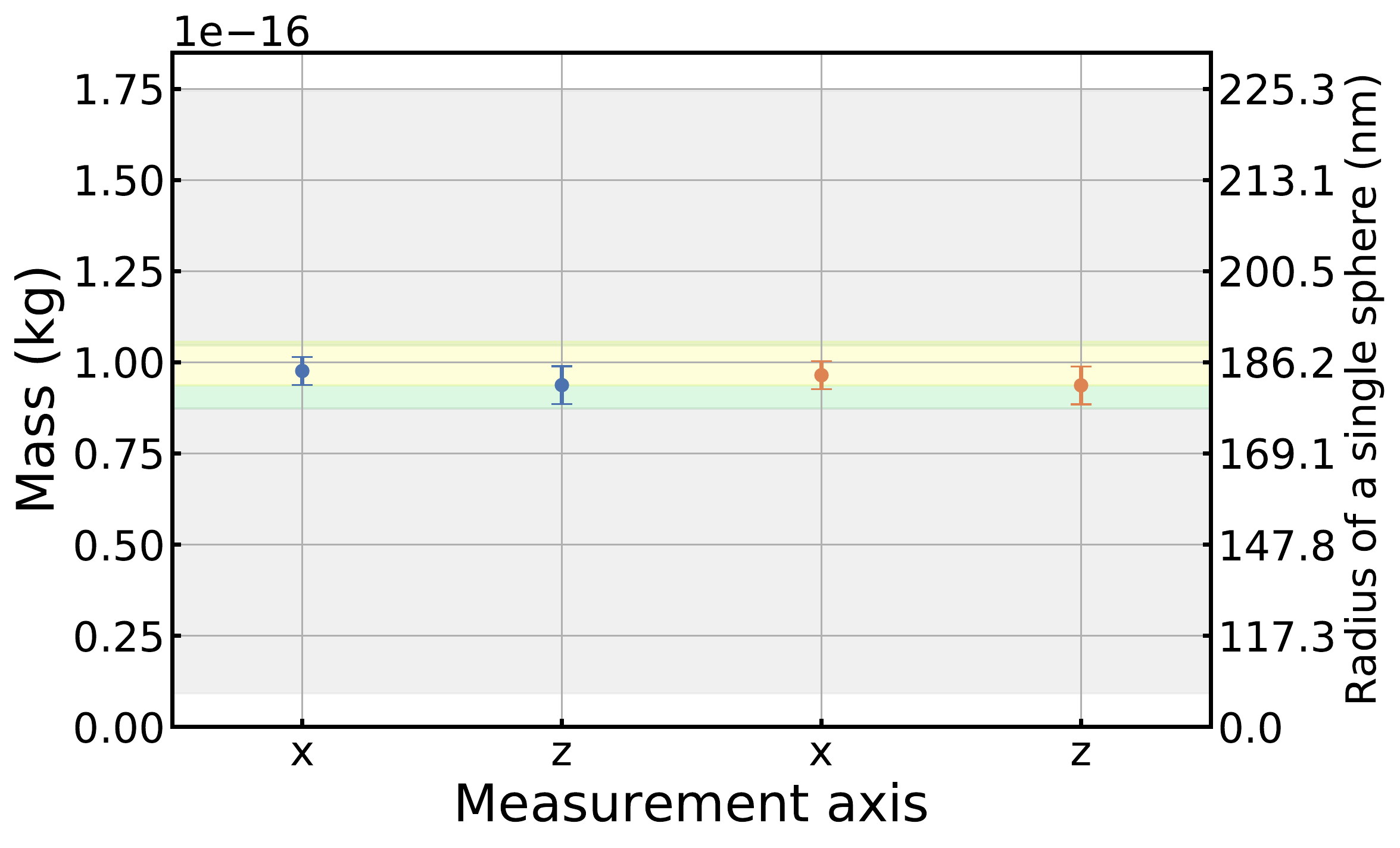}
\caption{The mass determination of the nanoparticle. The measurements in blue and orange are obtained by assuming thermal equilibrium of the motion at 293\,K with the displacement calibrated with the camera. The measurements are taken on two different days at a pressure of 1.0$\times$10$^{-2}$\,mbar. The calibration constant for the measurements in blue (orange) is (8.35$\,\pm\,0.08$)\,$\mu$m/pixel ((3.11$\,\pm\,0.03$)\,$\mu$m/pixel). The grey and green regions correspond to the uncertainty of the mass measurements (one standard deviation) with the linewidth measurement (see Fig.\,\ref{fig:size}\,(a)) and from the charge jumps (see Fig.\,\ref{fig:size}\,(b)), respectively. The yellow region corresponds to the mass estimated of two nanospheres from the SEM image, assuming a nominal density of 1850\,kg/m$^{3}$ (see Fig.\,\ref{fig1}\,(d)). The equivalent estimated averaged radius of the spheres forming the nanodumbbell is shown on the scale on the right-hand side, assuming a nominal density of 1850\,kg/m$^{3}$.}
\label{fig:radius}
\end{figure}

\bigskip
Fig.\,\ref{fig:radius} summarises the different measurements described above. The measurements in blue and orange correspond to the measured mass obtained from thermal equilibrium. Those measurements are taken at a pressure of 1.0$\times$10$^{-2}$\,mbar, with different camera magnifications and on two different days to check stability of the mass over time. The calibration used for the data in blue (orange) is (8.35$\,\pm\,0.08$)\,$\mu$m/pixel  ((3.11$\,\pm\,0.03$)\,$\mu$m/pixel). The green region corresponds to the uncertainty of the mass measurement (with one standard deviation) given by the charge jumps (see Fig.\,\ref{fig:size} (b)). As the particle size distribution, obtained from the SEM images, is narrow and that there is good agreement between the two mass measurements, we have good confidence that this nanoparticle is a nanodumbbell, which can easily form from two particles held by Van der Waals forces  \cite{nanodumbbell}. Indeed, assuming a nominal density of 1850\,kg/m$^{3}$ (provided by the manufacturer), a nanodumbbell made of spheres of radius ($186\pm4$)\,nm gives a mass of (9.9$\pm0.8)\times10^{-17}$\,kg (yellow region), which agrees very well with the different mass measurements. The density found when considering the mass given by the charge jumps with the size found on the SEM images gives (1781$\,\pm\,$196)\,kg/m$^{3}$. Lastly, the grey region corresponds to the expected mass given by the linewidth measurement assuming a cluster of two particles \cite{damping}. Those different measurements demonstrate the lack of reliability of the linewidth measurement in a Paul trap, as we have to guess the particle shape. Indeed, if we assume the particle to be spherical, we measure a radius of 199\,nm and if we assume a nanodumbbell made of two identical spheres, we estimate the radius (of one nanosphere) to be 182\,nm, which corresponds to a relative error in mass of 35\,\%. The equivalent radii for a single sphere of the nanodumbbell are shown in Fig.\,\ref{fig:radius}.

\subsection{Temperature estimation and stability}
\begin{figure}[H]
\centering
\begin{subfigure}{.5\textwidth}
  \centering
  \includegraphics[width=1\linewidth]{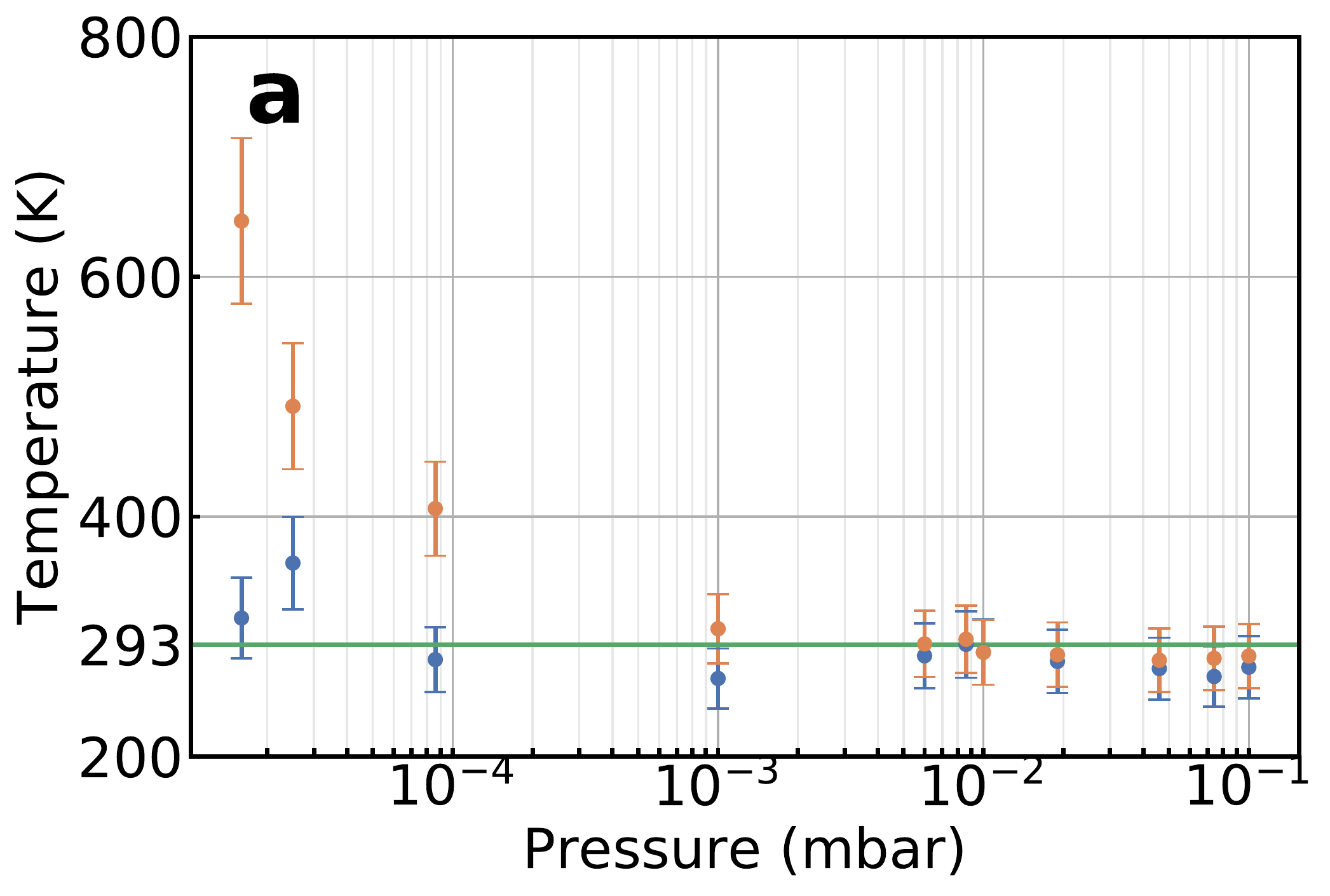}
  \caption{}
  \label{fig:sub1}
\end{subfigure}%
\begin{subfigure}{.5\textwidth}
\hspace{1.5mm}
  \centering
  \includegraphics[width=1\linewidth]{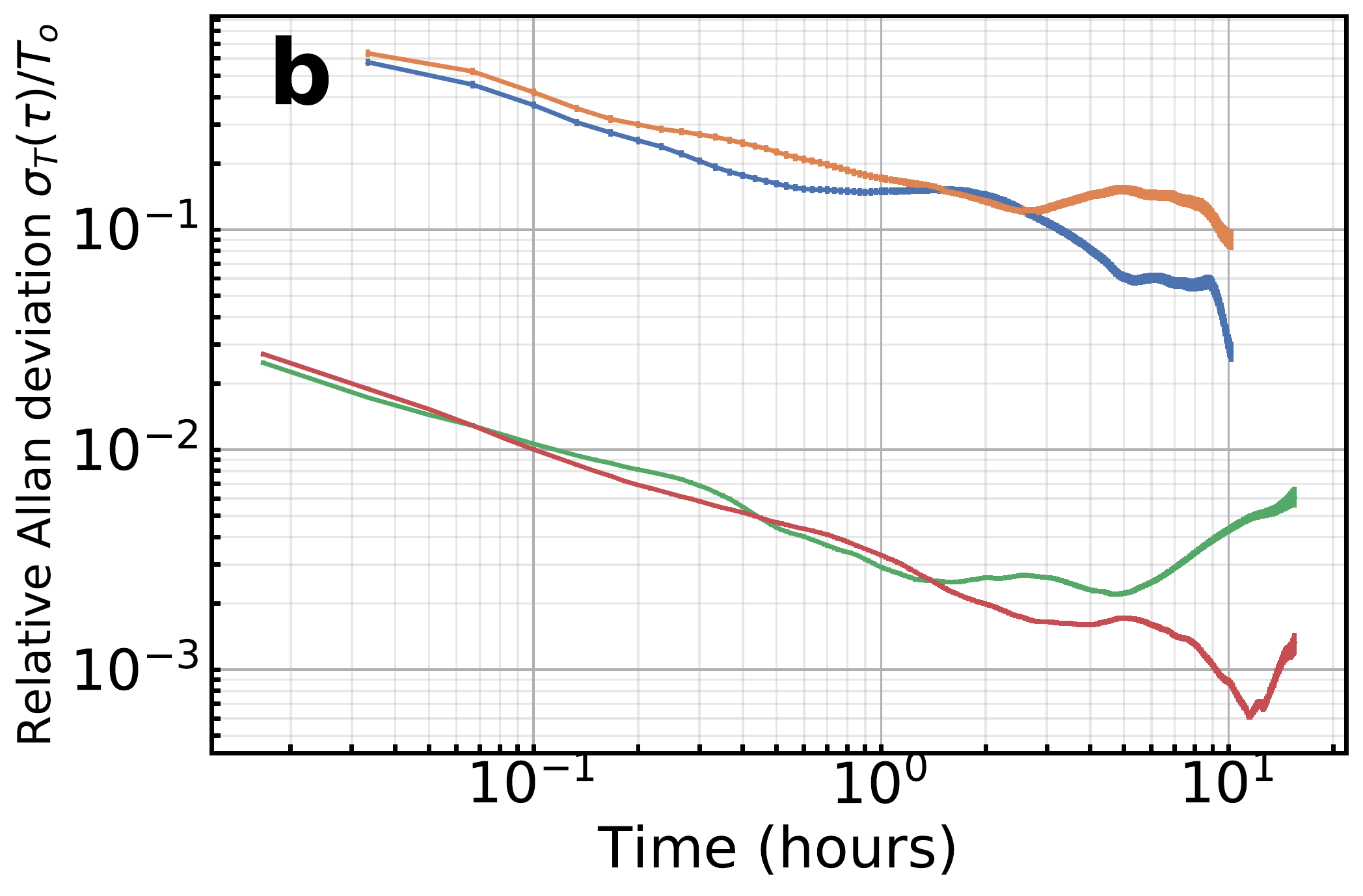}
  \caption{}
  \label{fig:sub2}
\end{subfigure}

\caption{(a) The trapped centre-of-mass temperature of the particle as a function of pressure (without cooling). The temperature along the \textit{x}-axis (\textit{z}-axis) is shown in blue (orange). The motion is thermal  ((282\,$\pm$\,25)\,K) down to $\sim10^{-4}$\,mbar along the \textit{z}-axis and to less than $\sim10^{-5}$\,mbar along the \textit{x}-axis. The temperature of the room at 293\,K is indicated in green. (b) The relative Allan deviation of the temperature at 1.6$\times$10$^{-5}$\,mbar along the \textit{x}-axis (\textit{z-}axis) is shown in blue (orange) and at 7.2$\times$10$^{-2}$\,mbar in green (red).}
\label{fig:temperature}
\end{figure}

We use the mass measurement from the charge jumps to estimate the centre-of-mass temperature. The temperature is estimated at different pressures without any active cooling mechanism on the particle. The temperature is calculated in Fig.\,\ref{fig:temperature}\,(a) with the motion calibrated with the camera. The blue (orange) data correspond to the motion along the \textit{x} (\textit{z}-axis). This proves that the motion is thermal ((282\,$\pm$\,25)\,K) down to $\sim10^{-4}$\,mbar in the \textit{z-}axis and to less than $\sim10^{-5}$\,mbar in the \textit{x-}axis. However, excess force noise increases the effective temperature of the centre-of-mass motion at lower pressures. Despite the increasing temperature at low pressures, the particle can be kept for weeks at 10$^{-7}$\,mbar.

We are also interested in the stability of the effective temperature over time, as well as the optimum time over which this measurement should be made. In Fig.\,\ref{fig:temperature}\,(b) we show the relative Allan deviation of the temperature $\sigma_{T}(\tau)/T_o$ at different pressures. We define the relative Allan deviation of a variable $V$ as $\sigma_{V}(\tau)/V_o$ with $V_o$ the averaged value of $V$ and $\sigma_{V}(\tau)$ the Allan deviation \cite{Allan1966}

\begin{equation}
\label{eq4}
\hspace{15mm}
\sigma_{V}^{2}(\tau)=\frac{1}{N-1}\sum_{k=1}^{N-1}\frac{1}{2}\left(\bar{V}_{k+1}^{(\tau)}-\bar{V}_{k}^{(\tau)}\right)\,,
\end{equation}

\noindent
where $\bar{V}_{k}^{(\tau)}$ corresponds to the time average value of $V$ inside $N$-1 intervals of varying length $\tau$ with $N$, such that $N\tau$ corresponds to the total measurement time. The Allan variance quantifies the stability of the system over time by effectively showing the optimal value of time needed to estimate the temperature. The measurements in blue (orange) are taken at 1.6$\times10^{-5}$\,mbar monitored for 23\,hours along the \textit{x}-axis (\textit{z-}axis). We also show measurements at 7.2$\times10^{-2}$\,mbar in green (red) along the \textit{x}-axis (\textit{z-}axis). The initial values and slopes match with the expected relative Allan deviation of a thermal oscillator $\sqrt{\frac{2}{\gamma \tau}}$\cite{Hebestreit2018}. This shows stability of the oscillator for a few hours, limited by the stability in the electrical potential. In comparison, stability over the 100\,seconds range has been reported in optical tweezers, which were believed to be limited by the optical stability of the trap as well as its nonlinearities \cite{Hebestreit2018,Ricci2019}. Similar stability of $\sim100$\,s were reported as well in other high frequency optomechanical systems \cite{Antonio}.

\bigskip
\section{Paul trap stability}
The particle stability in the Paul trap depends on several parameters such as the trapping mechanism, the pressure gauge and the material used for the trap itself.

\begin{figure}[!ht]
\centering
\begin{subfigure}{.5\textwidth}
  \centering
  \includegraphics[width=1\linewidth]{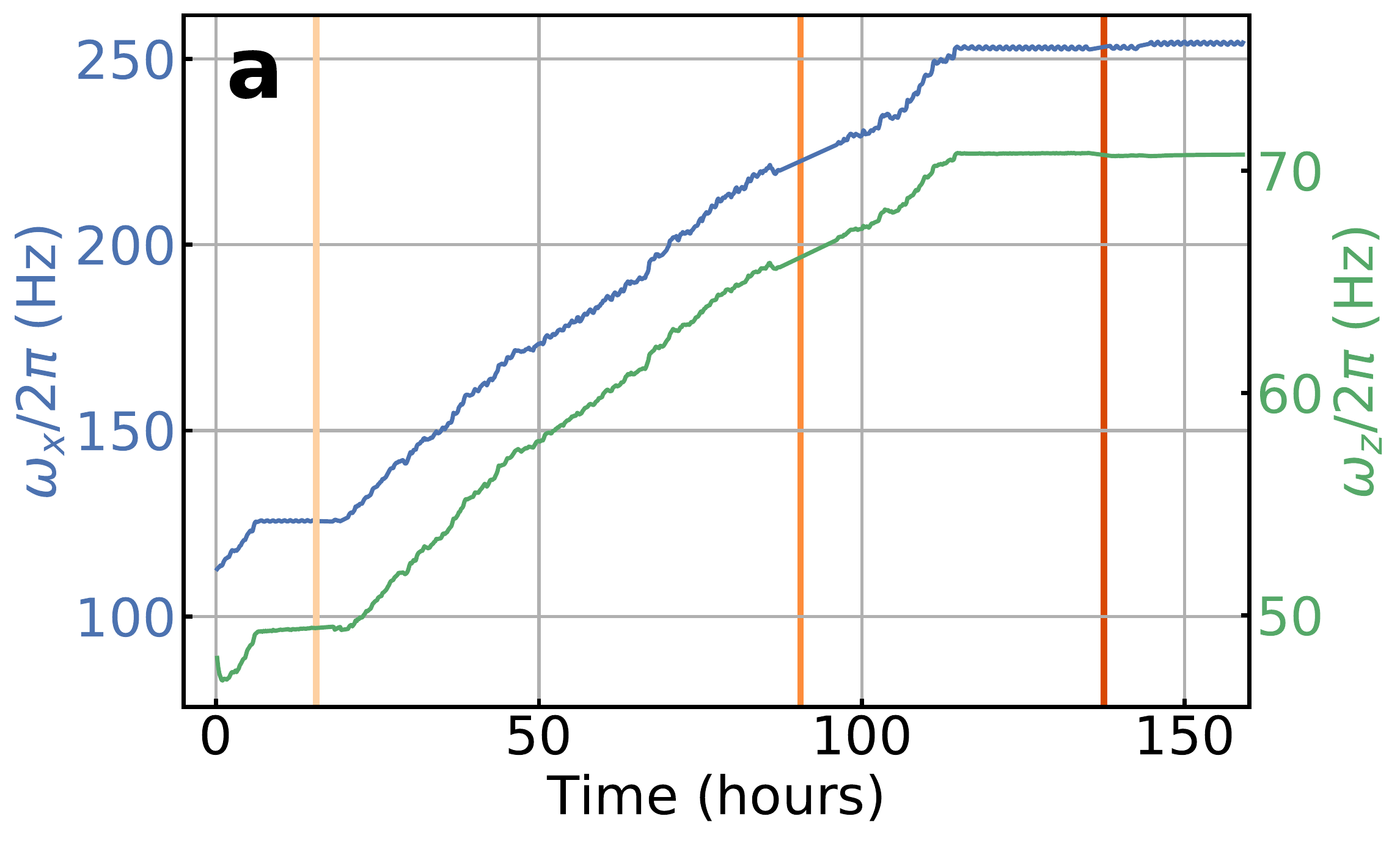}
  \caption{}
  \label{fig:sub1}
\end{subfigure}%
\begin{subfigure}{.5\textwidth}
  \centering
  \includegraphics[width=.9\linewidth]{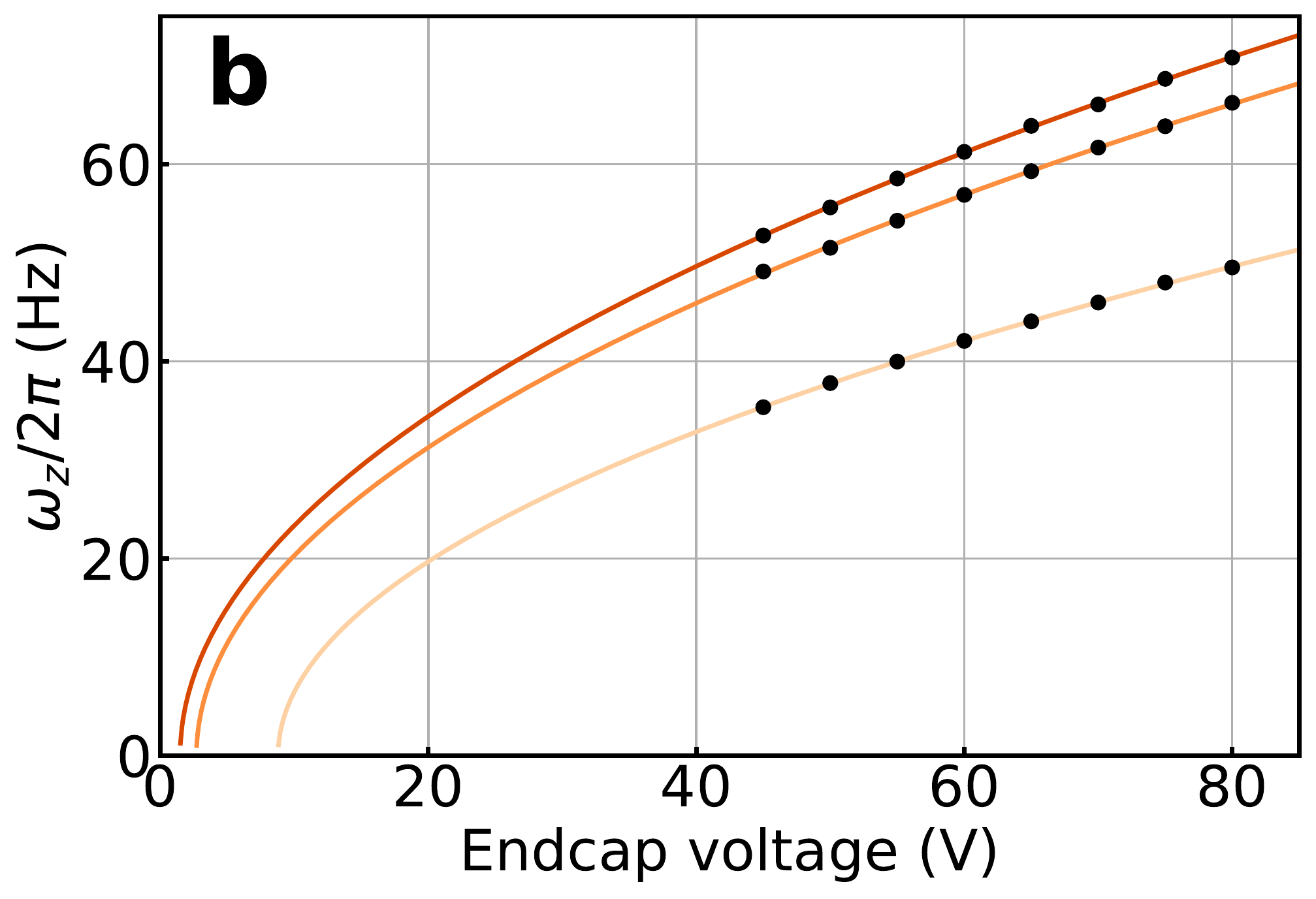}
  \caption{}
  \label{fig:sub2}
\end{subfigure}
\caption{(a) Measurements of secular frequencies monitored over 6 days. $\omega_x/2\pi$ is shown on the left ordinate (blue dataset), $\omega_z/2\pi$ on the right one (green dataset). The three colour bands mark the time at which the charge-to-mass ratios were measured and are shown in Fig.\,\ref{fig:stability} (b). (b) Fits of the charge-to-mass ratios at different times by changing the end-cap voltage and monitoring the secular motion along the \textit{z-}axis. The different times correspond to 15, 90 and 137\,hours, with charge-to-mass ratios of 0.10\,C/kg, 0.16\,C/Kg and 0.18\,C/kg, respectively, shown going from lighter to darker colours. The stray field is anti-trapping (negative effective potential for a positively charged particle) as it needs to be compensated with a positive voltage to trap the particle. The strength of this stray field decreases over time.}
\label{fig:stability}
\end{figure}

\subsection{Stray fields due to the ion gauge and the electrospray loading mechanism}

The creation of ions by the vacuum gauge leads to charging of the nanoparticle and changes in the trap potential. This leads to temporal changes in secular frequency as well as in the particle's mean position. The gauge therefore needs to be turned off for stable measurements. We monitor the secular frequencies $\omega_x$ and $\omega_z$ for almost one week when the gauge is turned on and off (see Fig.\,\ref{fig:stability} (a)). Both of these secular frequencies show very similar behaviour except initially, right after loading the trap. When the gauge is on, they follow an almost monotonic increase. The stable region shown in Fig.\,\ref{fig:stability} (b) occurs when the pressure gauge has been turned off. To some extent, it is possible to separate the change in number of charges from a change in the potential. To do this we measured the charge-to-mass ratio at different times ($t=15$, $90$ and $137$\, hours) by changing the endcap potential and measuring the resulting frequency shift. The results are shown in Fig.\,\ref{fig:stability}\,(a) and (b) along with the respective fits which give an increasing charge-to-mass ratio. We find $0.10$\,C/kg, $0.16$\,C/kg and  $0.18$\,C/kg ($10\%$ standard error), from bottom to top. However, an additional stray potential needs to be accounted for, in order to obtain a reasonable agreement with the data. This offset directly quantifies the stray field along the \textit{z}-axis as it corresponds to the minimum amount of endcap voltage needed to trap the particle which is ideally 0\,V (see Fig.\,\ref{fig:stability}\,(b)). Here, a positive voltage is needed (8.8, 2.7 and 1.5\,V, respectively).

Stray fields are also due to solvent droplets reaching the dielectric support of the trap during the loading phase. If the electrospray is kept in operation for sufficiently long times, the stray field can become quite strong even allowing for trapping in the \textit{z}-direction without any additional endcap potential and providing a trap frequency in this axis of the order of $\sim100$\,Hz. To verify this description we inverted the sign of the high voltage for the electrospray needle for a given time, to neutralise the charge build-up, to then go back to the usual configuration. As expected we found that in such a situation, the stray field given by the electrospray is anti-trapping; meaning more endcap potential is needed to trap the particle, as it is shown in Fig.\,\ref{fig:stability}\,(b). Furthermore, its strength is slowly decreasing as mentioned above.

\subsection{Potential drifts}

\begin{figure}[!ht]
\centering
    \begin{subfigure}{.5\textwidth}
        \centering
        
        \includegraphics[width=1\linewidth]{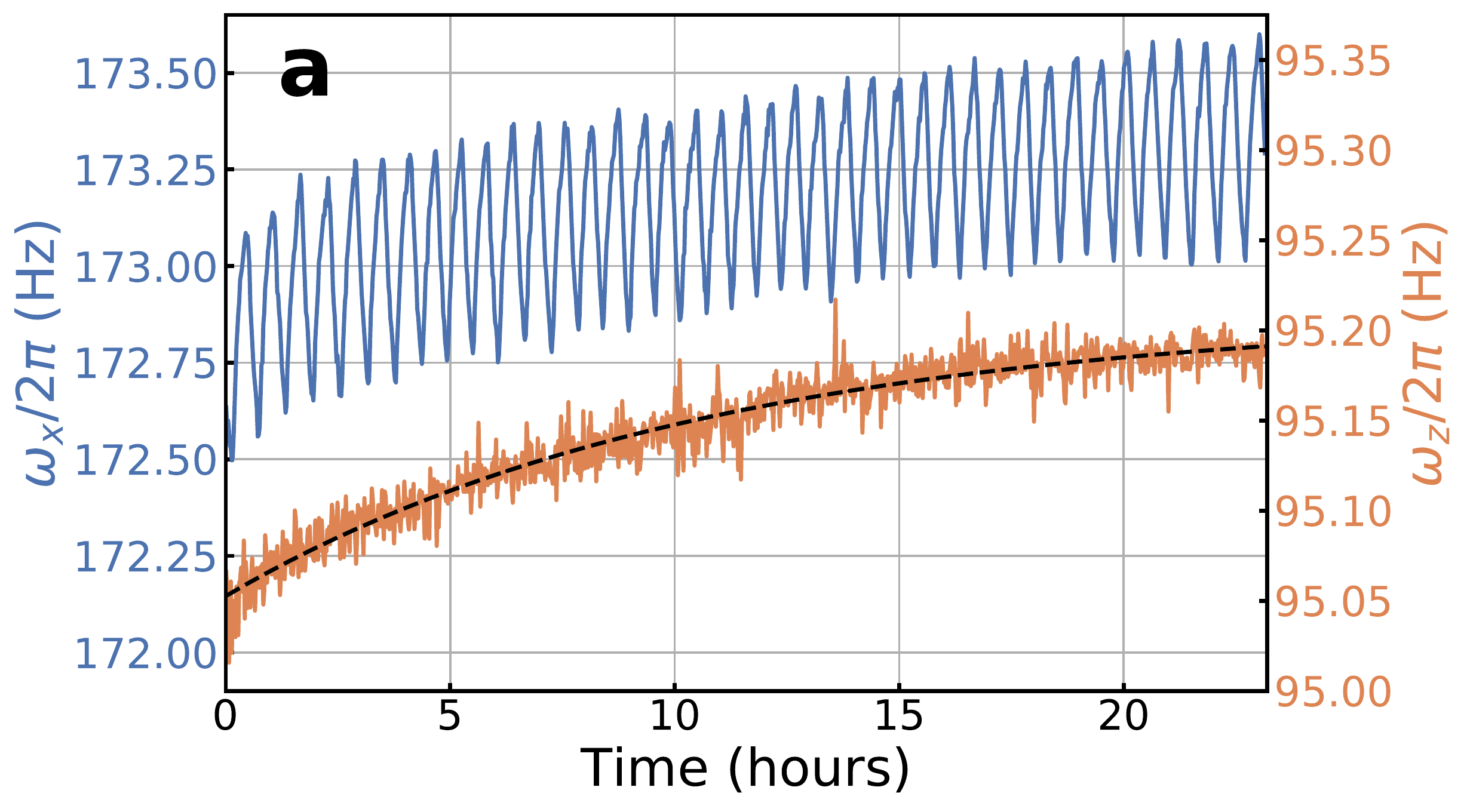}
        \caption{}
        \label{fig:sub1}
    \end{subfigure}
    \begin{subfigure}{.48\textwidth}
        \centering
        \includegraphics[width=1\linewidth]{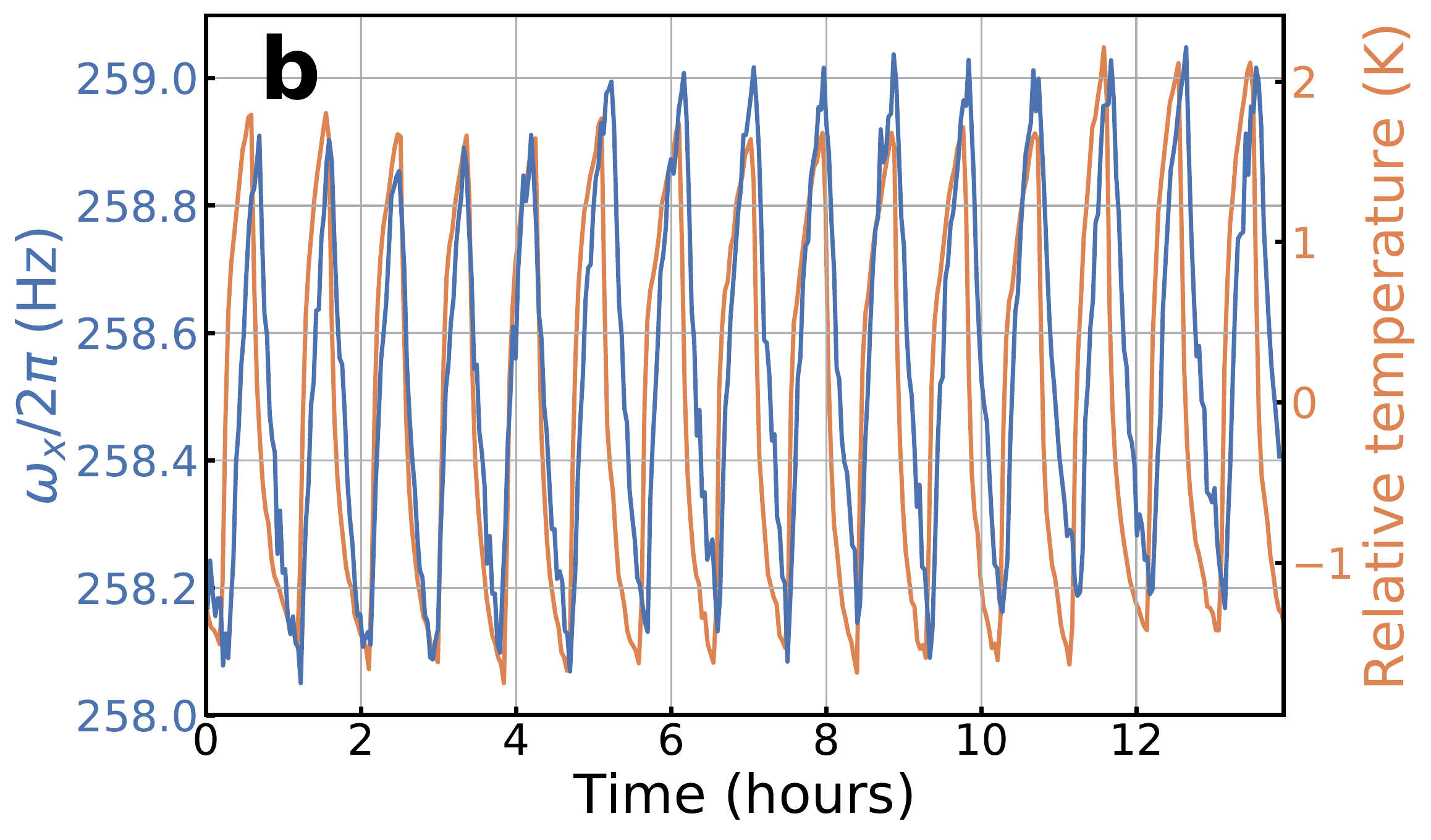}
         \caption{}
         \label{fig:sub2}
    \end{subfigure}
    \begin{subfigure}{.49\textwidth}
        \centering
        \includegraphics[width=1\linewidth]{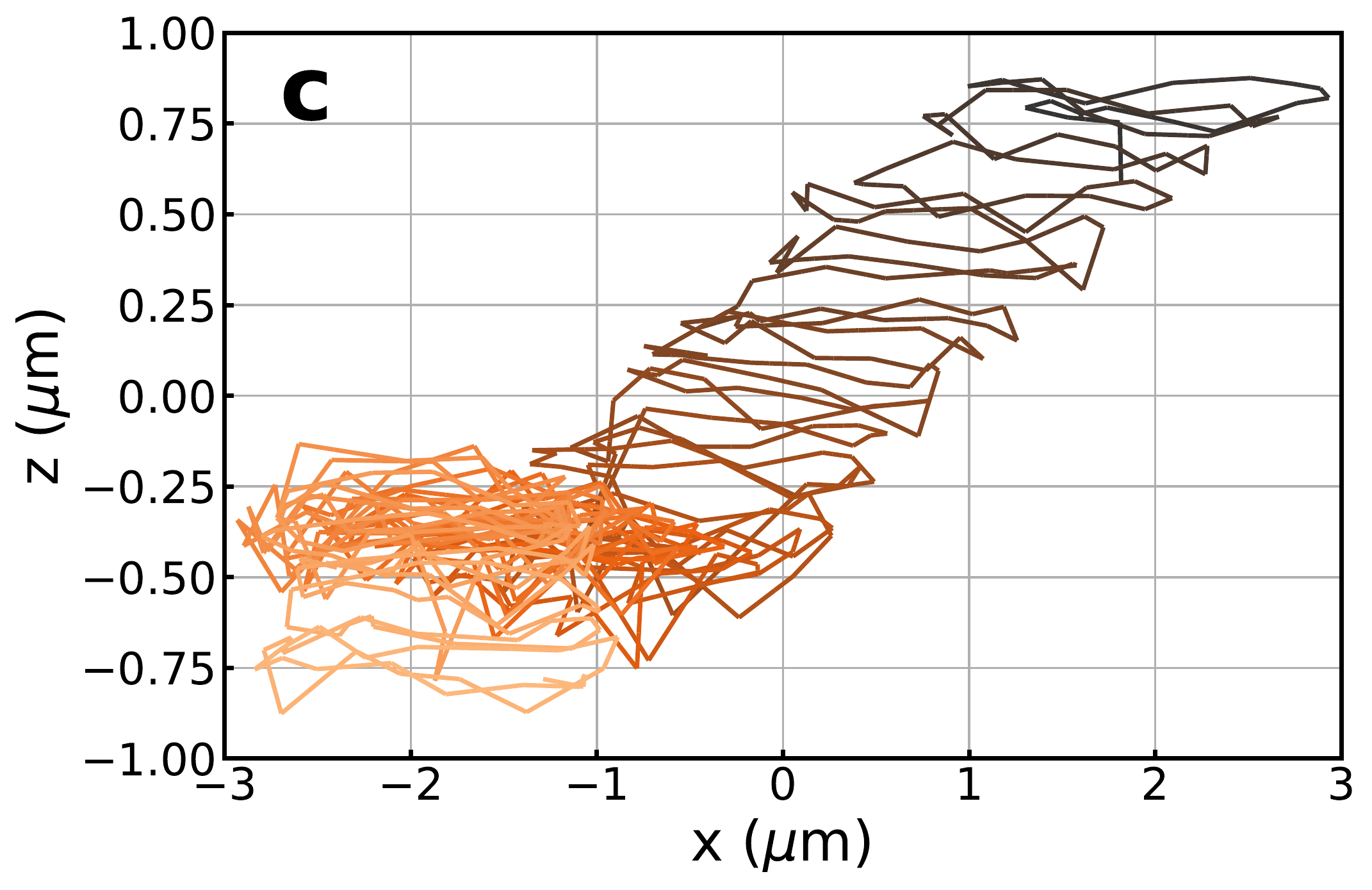}
        \caption{}
        \label{fig:sub2}
    \end{subfigure}
    \begin{subfigure}{.49\textwidth}
        \centering
        \includegraphics[width=1\linewidth]{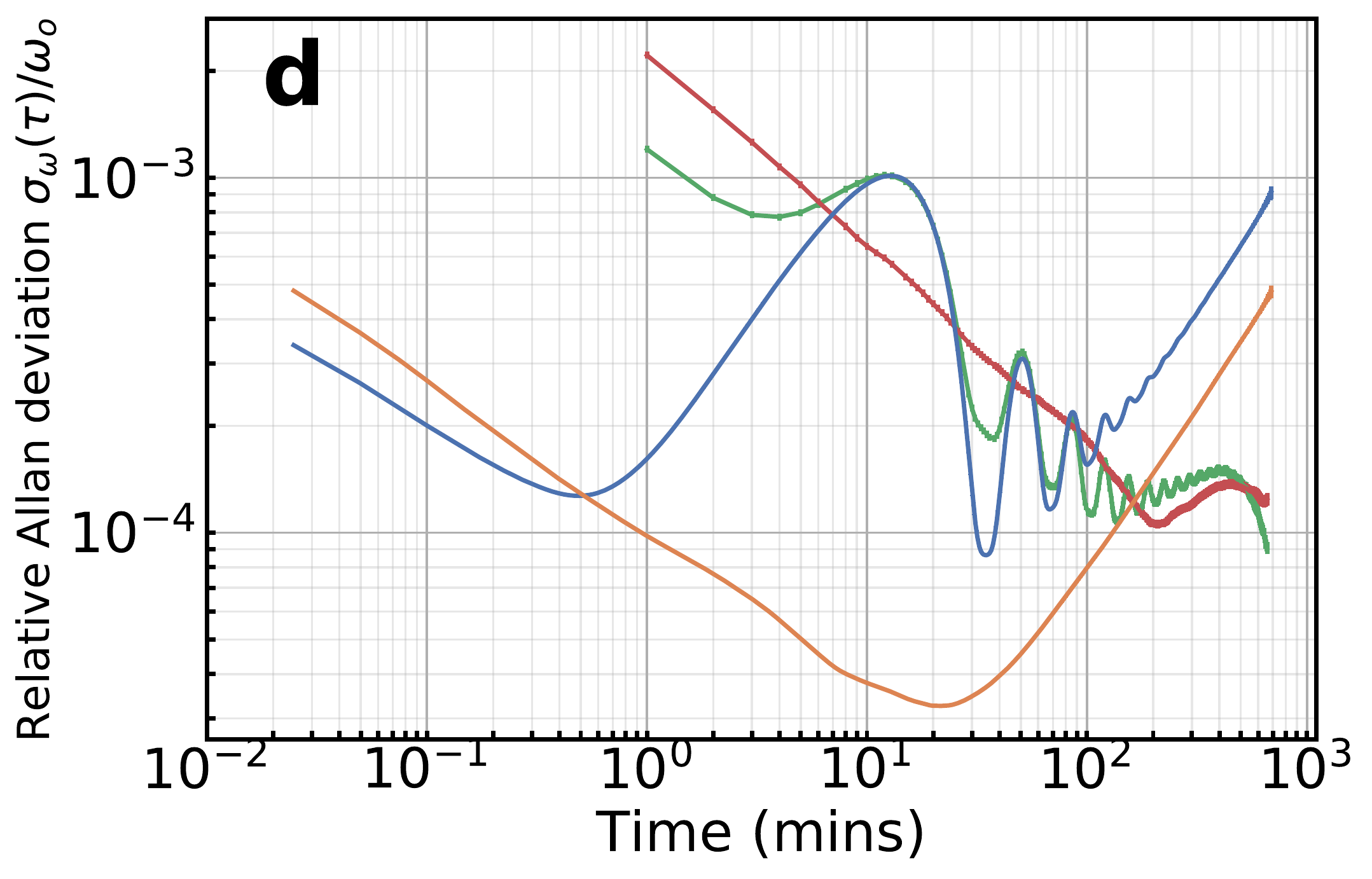}
        \caption{}
        \label{fig:sub2}
    \end{subfigure}
    \caption{(a) Measurement of the secular frequency over time along the \textit{x} (\textit{z}-axis) is shown in blue (orange) while keeping the particle at a pressure of $1.6\times10^{-5}$\,mbar. The measurements are taken right after turning the pressure gauge off. An exponential rise of 10.7\,hours is fitted along the \textit{z}-axis (fit shown in black dotted lines). (b) A plot of the temperature drifts of the room (in orange) and of the secular frequency (in blue) along the \textit{x}-axis at the same time. (c) Measurements of the drifts of the particle mean position over time (from dark colours to light ones) in the trap along the \textit{x} and \textit{z}-directions for the same data set as the one shown in (a). (d) The relative Allan deviation of the secular frequency along the \textit{x}-axis (\textit{z}-axis) is shown in blue (orange) at $1.6\times10^{-5}$\,mbar and in green (red) at $7.6\times10^{-2}$\,mbar. \textit{Note:} The data shown in (a), (c) and (d) at $1.6\times10^{-5}$\,mbar are from the same data set with the same particle as the one characterised in Section \ref{sec:particle}.}
    \label{fig:drifts}
\end{figure}

Right after turning the ion gauge off, we have observed an exponential decay in the trapping field with a time constant of approximately 10\,hours. This characteristic time, consistently measured over tens of traces is likely to be due to the dielectric material used. Dielectric surfaces should therefore be kept in any trap as far as possible from the nanoparticle. Some of the charging effects during the loading phase could be mitigated by using both a bent and longer guide, or by using a loading mechanism free of solvent \cite{Schlemmer,Tracy}. In Fig.\,\ref{fig:drifts}\,(a), we show the stability of the secular frequencies along the \textit{x} and \textit{z}-axis right after turning the pressure gauge off. The exponential rise along the \textit{z}-axis has a characteristic time of 10.7\,hours. On top of the exponential rise, a very periodic modulation (period of $\sim$20\,mins) can be seen in the secular frequency along the \textit{x}-axis. This modulation is correlated to changes in the temperature of the room as shown in Fig.\,\ref{fig:drifts}\,(b), and is fully attributable to temperature induced changes in the amplitude of the signal generator, high voltage amplifier and other electronics used to provide the trapping AC field. This corresponds to a slow modulation of approximately 1\,V (peak to peak) in the amplitude of the signal applied to the trap electrodes (0.2\% of the applied signal). Those drifts could easily be reduced with a better temperature stabilisation.

In Fig.\,\ref{fig:drifts} (c), we show the averaged position in the \textit{x-z} plane over time for the same data as shown in Fig.\,\ref{fig:drifts} (a). One sample corresponds to 2.5\,mins. The time is shown with a colour gradient, starting with black at $t=0$, and ending in yellow at $t=23$\,hours. Correlations can be seen between the drifts in the secular frequency along the \textit{z}-axis and the motion along the same axis. The same can be noticed along the \textit{x}-axis. This shows that the same potential drifts are responsible for both changes in the mean position of the nanoparticle in the trap as well as its secular frequencies. Lastly, we show in Fig.\,\ref{fig:drifts} (d) the relative Allan deviation $\sigma_{f}(\tau)/f_o$ of the two secular frequencies monitored for 23\,hours with $f_o$ the averaged frequency. The Allan deviation is calculated on the frequency time traces shown in Fig.\,\ref{fig:drifts}\,(a). We get in this case a frequency stability of more than 20\,min with a relative uncertainty on the frequency of $\sim$30\,ppm along the \textit{z}-axis (orange line) and with overall drifts of 60\,ppm/h. Along the \textit{x}-axis (blue line), we recover the periodicity ($\sim$30\,mins) also shown in Fig.\,\ref{fig:drifts}\,(a), as the frequency periodically gets closer to its initial value. We show in Fig.\,\ref{fig:drifts}\,(a) the stability of the frequency along the \textit{z}-axis (\textit{x}-axis) at 7.2$\times10^{-2}$\,mbar. These data cover 32\,hours of continuous acquisition that started 33\,hours after having turned off the pressure gauge, to ensure a better stability. As expected, longer optimal times of at least 5\,hours were obtained. The optimum time is likely to be even longer since the drifts are not yet limiting the Allan deviation. Furthermore the behaviour above 500\,mins is likely to be caused by aliasing of the frequency fluctuations. Lastly, in this more stable case, a low frequency drift of 2ppm/h (21ppm/h) was obtained along the \textit{x}-axis (\textit{z}-axis). This demonstrates, that despite the use of a dielectric and the different stray fields mentioned above, a competitive frequency stability can still be achieved \cite{Schlemmer}.   


\section{Conclusion}
To conclude, we have characterised the size, temperature and stability of a charged particle levitated nano-oscillator. Thermal equilibrium and charge jumps can be used to measure efficiently the mass of a nanoparticle in a Paul trap, providing sufficient accuracy to distinguish between masses of one or several particles. In fact, we clearly demonstrate trapping of a nanodumbbell. As shown here, when the oscillator frequency is small enough ($<$ 1 kHz), a camera can be used to calibrate the motion accurately. We have shown that the PCB substrate, loading mechanism, temperature fluctuations and ion gauge can modify the trapping potential and therefore the stability of the oscillator. Despite those fluctuations, drifts as small as 2\,ppm/h in the trapping frequency were obtained over hours of measurement. Furthermore, a temperature stability larger than five hours were reported with a stable number of charges over weeks. The stray fields discussed above can be mitigated with stronger filtering of the charge-to-mass ratios of nanoparticles in the guide in combination with a longer and bent guide. Furthermore, the internal face of the PCB board can be fully grounded \cite{Brown} to considerably reduce the stray fields accumulated on the dielectric surface due to both the loading process and the ion gauge. Those changes should increase the overall stability of the system needed for experiments which aim to use nanoparticles in a Paul trap for tests of wavefunction collapse \cite{Goldwater,TEQ}. In addition, the precise knowledge of the mass and number of charges, combined with the methods identified to increase stability will significantly enhance experiments with the hybrid electro-optical trap which aims to cool trapped particles to their ground state \cite{Fonseca2016}. Lastly, the ability to stably trap and characterise the centre-of-mass motion of an anisotropic particle like a nanodumbbell will be useful for future experiments that aim to explore rotational optomechanics in the absence of optical fields \cite{Stickler2018}.

\section*{Acknowledgements}
The authors would like to thank Thomas Penny for interesting discussions and Jonathan Gosling for taking the SEM picture. The authors acknowledge funding from the EPSRC Grant No. EP/N031105/1 and from the EU H2020 FET project TEQ (Grant No. 766900). NPB acknowledges funding from the EPSRC Grant No. EP/L015242/1. AP has received funding from the European Union’s Horizon 2020 research and innovation programme under the Marie Sklodowska-Curie Grant Agreement No. 749709.

\bigskip
\bigskip

\bibliographystyle{nicebib}

\section*{References}
\bibliography{trap_bib}
\end{document}